\documentclass[twocolumn]{article}

\usepackage[letterpaper,bottom=17.4mm,top=10mm,left=17mm,right=17.9mm,columnsep=5mm,headsep=0mm,headheight=0mm,footskip=18mm]{geometry}

\usepackage[T1]{fontenc}

\usepackage{times}

\usepackage{mleftright}
\usepackage{cite}
\usepackage{graphicx}
\usepackage{xcolor}
\usepackage{mathtools}
\usepackage{amssymb}
\usepackage{amsmath}
\usepackage{paralist}
\usepackage{subfigure}
\usepackage{booktabs} 
\usepackage{algorithm}
\usepackage{algorithmic}
\usepackage{amsthm}
\usepackage{multirow}
\usepackage{microtype}
\usepackage{tablefootnote}
\usepackage{stfloats}
\usepackage{caption}
\usepackage{circledsteps}
\usepackage{tikz}
\usepackage{xfrac}
\usepackage{array}
\usepackage{eso-pic}
\newcolumntype{P}[1]{>{\centering\arraybackslash}p{#1}}
\usepackage{amssymb}
\usepackage{amsfonts}
\usepackage{mathrsfs}
\usepackage{xspace}
\usepackage{bm}
\usepackage{upgreek}

\newcommand{\safemath}[2]{\newcommand{#1}{\ensuremath{#2}\xspace}}

\safemath{\bma}{\mathbf{a}}
\safemath{\bmb}{\mathbf{b}}
\safemath{\bmc}{\mathbf{c}}
\safemath{\bmd}{\mathbf{d}}
\safemath{\bme}{\mathbf{e}}
\safemath{\bmf}{\mathbf{f}}
\safemath{\bmg}{\mathbf{g}}
\safemath{\bmh}{\mathbf{h}}
\safemath{\bmi}{\mathbf{i}}
\safemath{\bmj}{\mathbf{j}}
\safemath{\bmk}{\mathbf{k}}
\safemath{\bml}{\mathbf{l}}
\safemath{\bmm}{\mathbf{m}}
\safemath{\bmn}{\mathbf{n}}
\safemath{\bmo}{\mathbf{o}}
\safemath{\bmp}{\mathbf{p}}
\safemath{\bmq}{\mathbf{q}}
\safemath{\bmr}{\mathbf{r}}
\safemath{\bms}{\mathbf{s}}
\safemath{\bmt}{\mathbf{t}}
\safemath{\bmu}{\mathbf{u}}
\safemath{\bmv}{\mathbf{v}}
\safemath{\bmw}{\mathbf{w}}
\safemath{\bmx}{\mathbf{x}}
\safemath{\bmy}{\mathbf{y}}
\safemath{\bmz}{\mathbf{z}}
\safemath{\bmzero}{\mathbf{0}}
\safemath{\bmone}{\mathbf{1}}

\bmdefine{\biad}{a}
\bmdefine{\bibd}{b}
\bmdefine{\bicd}{c}
\bmdefine{\bidd}{d}
\bmdefine{\bied}{e}
\bmdefine{\bifd}{f}
\bmdefine{\bigd}{g}
\bmdefine{\bihd}{h}
\bmdefine{\biid}{i}
\bmdefine{\bijd}{j}
\bmdefine{\bikd}{k}
\bmdefine{\bild}{l}
\bmdefine{\bimd}{m}
\bmdefine{\bind}{n}
\bmdefine{\biod}{o}
\bmdefine{\bipd}{p}
\bmdefine{\biqd}{q}
\bmdefine{\bird}{r}
\bmdefine{\bisd}{s}
\bmdefine{\bitd}{t}
\bmdefine{\biud}{u}
\bmdefine{\bivd}{v}
\bmdefine{\biwd}{w}
\bmdefine{\bixd}{x}
\bmdefine{\biyd}{y}
\bmdefine{\bizd}{z}

\bmdefine{\bixid}{\xi}
\bmdefine{\bilambdad}{\lambda}
\bmdefine{\bimud}{\mu}
\bmdefine{\bithetad}{\theta}
\bmdefine{\biphid}{\phi}
\bmdefine{\bideltad}{\delta}

\safemath{\bmia}{\biad}
\safemath{\bmib}{\bibd}
\safemath{\bmic}{\bicd}
\safemath{\bmid}{\bidd}
\safemath{\bmie}{\bied}
\safemath{\bmif}{\bifd}
\safemath{\bmig}{\bigd}
\safemath{\bmih}{\bihd}
\safemath{\bmii}{\biid}
\safemath{\bmij}{\bijd}
\safemath{\bmik}{\bikd}
\safemath{\bmil}{\bild}
\safemath{\bmim}{\bimd}
\safemath{\bmin}{\bind}
\safemath{\bmio}{\biod}
\safemath{\bmip}{\bipd}
\safemath{\bmiq}{\biqd}
\safemath{\bmir}{\bird}
\safemath{\bmis}{\bisd}
\safemath{\bmit}{\bitd}
\safemath{\bmiu}{\biud}
\safemath{\bmiv}{\bivd}
\safemath{\bmiw}{\biwd}
\safemath{\bmix}{\bixd}
\safemath{\bmiy}{\biyd}
\safemath{\bmiz}{\bizd}

\safemath{\bmxi}{\bixid}
\safemath{\bmlambda}{\bilambdad}
\safemath{\bmmu}{\bimud}
\safemath{\bmtheta}{\bithetad}
\safemath{\bmphi}{\biphid}
\safemath{\bmdelta}{\bideltad}

\safemath{\bA}{\mathbf{A}}
\safemath{\bB}{\mathbf{B}}
\safemath{\bC}{\mathbf{C}}
\safemath{\bD}{\mathbf{D}}
\safemath{\bE}{\mathbf{E}}
\safemath{\bF}{\mathbf{F}}
\safemath{\bG}{\mathbf{G}}
\safemath{\bH}{\mathbf{H}}
\safemath{\bI}{\mathbf{I}}
\safemath{\bJ}{\mathbf{J}}
\safemath{\bK}{\mathbf{K}}
\safemath{\bL}{\mathbf{L}}
\safemath{\bM}{\mathbf{M}}
\safemath{\bN}{\mathbf{N}}
\safemath{\bO}{\mathbf{O}}
\safemath{\bP}{\mathbf{P}}
\safemath{\bQ}{\mathbf{Q}}
\safemath{\bR}{\mathbf{R}}
\safemath{\bS}{\mathbf{S}}
\safemath{\bT}{\mathbf{T}}
\safemath{\bU}{\mathbf{U}}
\safemath{\bV}{\mathbf{V}}
\safemath{\bW}{\mathbf{W}}
\safemath{\bX}{\mathbf{X}}
\safemath{\bY}{\mathbf{Y}}
\safemath{\bZ}{\mathbf{Z}}

\safemath{\bZero}{\mathbf{0}}
\safemath{\bOne}{\mathbf{1}}
\safemath{\bDelta}{\mathbf{\Delta}}
\safemath{\bLambda}{\mathbf{\UpLambda}}
\safemath{\bPhi}{\mathbf{\Upphi}}
\safemath{\bSigma}{\mathbf{\Upsigma}}
\safemath{\bOmega}{\mathbf{\Upomega}}
\safemath{\bTheta}{\mathbf{\Uptheta}}

\bmdefine{\biAd}{A}
\bmdefine{\biBd}{B}
\bmdefine{\biCd}{C}
\bmdefine{\biDd}{D}
\bmdefine{\biEd}{E}
\bmdefine{\biFd}{F}
\bmdefine{\biGd}{G}
\bmdefine{\biHd}{H}
\bmdefine{\biId}{I}
\bmdefine{\biJd}{J}
\bmdefine{\biKd}{K}
\bmdefine{\biLd}{L}
\bmdefine{\biMd}{M}
\bmdefine{\biOd}{N}
\bmdefine{\biPd}{O}
\bmdefine{\biQd}{P}
\bmdefine{\biRd}{R}
\bmdefine{\biSd}{S}
\bmdefine{\biTd}{T}
\bmdefine{\biUd}{U}
\bmdefine{\biVd}{V}
\bmdefine{\biWd}{W}
\bmdefine{\biXd}{X}
\bmdefine{\biYd}{Y}
\bmdefine{\biZd}{Z}

\bmdefine{\biDelta}{\Delta}
\bmdefine{\biLambda}{\Lambda}
\bmdefine{\biPhi}{\Phi}
\bmdefine{\biSigma}{\Sigma}
\bmdefine{\biOmega}{\Omega}
\bmdefine{\biTheta}{\Theta}

\safemath{\bimA}{\biAd}
\safemath{\bimB}{\biBd}
\safemath{\bimC}{\biCd}
\safemath{\bimD}{\biDd}
\safemath{\bimE}{\biEd}
\safemath{\bimF}{\biFd}
\safemath{\bimG}{\biGd}
\safemath{\bimH}{\biHd}
\safemath{\bimI}{\biId}
\safemath{\bimJ}{\biJd}
\safemath{\bimK}{\biKd}
\safemath{\bimL}{\biLd}
\safemath{\bimM}{\biMd}
\safemath{\bimN}{\biNd}
\safemath{\bimO}{\biOd}
\safemath{\bimP}{\biPd}
\safemath{\bimQ}{\biQd}
\safemath{\bimR}{\biRd}
\safemath{\bimS}{\biSd}
\safemath{\bimT}{\biTd}
\safemath{\bimU}{\biUd}
\safemath{\bimV}{\biVd}
\safemath{\bimW}{\biWd}
\safemath{\bimX}{\biXd}
\safemath{\bimY}{\biYd}
\safemath{\bimZ}{\biZd}

\safemath{\bimDelta}{\biDelta}
\safemath{\bimLambda}{\biLambda}
\safemath{\bimPhi}{\biPhi}
\safemath{\bimSigma}{\biSigma}
\safemath{\bimOmega}{\biOmega}
\safemath{\bimTheta}{\biTheta}

\safemath{\setA}{\mathcal{A}}
\safemath{\setB}{\mathcal{B}}
\safemath{\setC}{\mathcal{C}}
\safemath{\setD}{\mathcal{D}}
\safemath{\setE}{\mathcal{E}}
\safemath{\setF}{\mathcal{F}}
\safemath{\setG}{\mathcal{G}}
\safemath{\setH}{\mathcal{H}}
\safemath{\setI}{\mathcal{I}}
\safemath{\setJ}{\mathcal{J}}
\safemath{\setK}{\mathcal{K}}
\safemath{\setL}{\mathcal{L}}
\safemath{\setM}{\mathcal{M}}
\safemath{\setN}{\mathcal{N}}
\safemath{\setO}{\mathcal{O}}
\safemath{\setP}{\mathcal{P}}
\safemath{\setQ}{\mathcal{Q}}
\safemath{\setR}{\mathcal{R}}
\safemath{\setS}{\mathcal{S}}
\safemath{\setT}{\mathcal{T}}
\safemath{\setU}{\mathcal{U}}
\safemath{\setV}{\mathcal{V}}
\safemath{\setW}{\mathcal{W}}
\safemath{\setX}{\mathcal{X}}
\safemath{\setY}{\mathcal{Y}}
\safemath{\setZ}{\mathcal{Z}}
\safemath{\emptySet}{\varnothing}

\safemath{\colA}{\mathscr{A}}
\safemath{\colB}{\mathscr{B}}
\safemath{\colC}{\mathscr{C}}
\safemath{\colD}{\mathscr{D}}
\safemath{\colE}{\mathscr{E}}
\safemath{\colF}{\mathscr{F}}
\safemath{\colG}{\mathscr{G}}
\safemath{\colH}{\mathscr{H}}
\safemath{\colI}{\mathscr{I}}
\safemath{\colJ}{\mathscr{J}}
\safemath{\colK}{\mathscr{K}}
\safemath{\colL}{\mathscr{L}}
\safemath{\colM}{\mathscr{M}}
\safemath{\colN}{\mathscr{N}}
\safemath{\colO}{\mathscr{O}}
\safemath{\colP}{\mathscr{P}}
\safemath{\colQ}{\mathscr{Q}}
\safemath{\colR}{\mathscr{R}}
\safemath{\colS}{\mathscr{S}}
\safemath{\colT}{\mathscr{T}}
\safemath{\colU}{\mathscr{U}}
\safemath{\colV}{\mathscr{V}}
\safemath{\colW}{\mathscr{W}}
\safemath{\colX}{\mathscr{X}}
\safemath{\colY}{\mathscr{Y}}
\safemath{\colZ}{\mathscr{Z}}

\safemath{\opA}{\mathbb{A}}
\safemath{\opB}{\mathbb{B}}
\safemath{\opC}{\mathbb{C}}
\safemath{\opD}{\mathbb{D}}
\safemath{\opE}{\mathbb{E}}
\safemath{\opF}{\mathbb{F}}
\safemath{\opG}{\mathbb{G}}
\safemath{\opH}{\mathbb{H}}
\safemath{\opI}{\mathbb{I}}
\safemath{\opJ}{\mathbb{J}}
\safemath{\opK}{\mathbb{K}}
\safemath{\opL}{\mathbb{L}}
\safemath{\opM}{\mathbb{M}}
\safemath{\opN}{\mathbb{N}}
\safemath{\opO}{\mathbb{O}}
\safemath{\opP}{\mathbb{P}}
\safemath{\opQ}{\mathbb{Q}}
\safemath{\opR}{\mathbb{R}}
\safemath{\opS}{\mathbb{S}}
\safemath{\opT}{\mathbb{T}}
\safemath{\opU}{\mathbb{U}}
\safemath{\opV}{\mathbb{V}}
\safemath{\opW}{\mathbb{W}}
\safemath{\opX}{\mathbb{X}}
\safemath{\opY}{\mathbb{Y}}
\safemath{\opZ}{\mathbb{Z}}
\safemath{\opZero}{\mathbb{O}}
\safemath{\identityop}{\opI}

\safemath{\veca}{\bma}
\safemath{\vecb}{\bmb}
\safemath{\vecc}{\bmc}
\safemath{\vecd}{\bmd}
\safemath{\vece}{\bme}
\safemath{\vecf}{\bmf}
\safemath{\vecg}{\bmg}
\safemath{\vech}{\bmh}
\safemath{\veci}{\bmi}
\safemath{\vecj}{\bmj}
\safemath{\veck}{\bmk}
\safemath{\vecl}{\bml}
\safemath{\vecm}{\bmm}
\safemath{\vecn}{\bmn}
\safemath{\veco}{\bmo}
\safemath{\vecp}{\bmp}
\safemath{\vecq}{\bmq}
\safemath{\vecr}{\bmr}
\safemath{\vecs}{\bms}
\safemath{\vect}{\bmt}
\safemath{\vecu}{\bmu}
\safemath{\vecv}{\bmv}
\safemath{\vecw}{\bmw}
\safemath{\vecx}{\bmx}
\safemath{\vecy}{\bmy}
\safemath{\vecz}{\bmz}

\safemath{\veczero}{\bmzero}
\safemath{\vecone}{\bmone}
\safemath{\vecxi}{\bmxi}
\safemath{\veclambda}{\bmlambda}
\safemath{\vecmu}{\bmmu}
\safemath{\vectheta}{\bmtheta}
\safemath{\vecphi}{\bmphi}
\safemath{\vecdelta}{\bmdelta}

\safemath{\matA}{\bA}
\safemath{\matB}{\bB}
\safemath{\matC}{\bC}
\safemath{\matD}{\bD}
\safemath{\matE}{\bE}
\safemath{\matF}{\bF}
\safemath{\matG}{\bG}
\safemath{\matH}{\bH}
\safemath{\matI}{\bI}
\safemath{\matJ}{\bJ}
\safemath{\matK}{\bK}
\safemath{\matL}{\bL}
\safemath{\matM}{\bM}
\safemath{\matN}{\bN}
\safemath{\matO}{\bO}
\safemath{\matP}{\bP}
\safemath{\matQ}{\bQ}
\safemath{\matR}{\bR}
\safemath{\matS}{\bS}
\safemath{\matT}{\bT}
\safemath{\matU}{\bU}
\safemath{\matV}{\bV}
\safemath{\matW}{\bW}
\safemath{\matX}{\bX}
\safemath{\matY}{\bY}
\safemath{\matZ}{\bZ}
\safemath{\matzero}{\bmzero}

\safemath{\matDelta}{\bDelta}
\safemath{\matLambda}{\bLambda}
\safemath{\matPhi}{\bPhi}
\safemath{\matSigma}{\bSigma}
\safemath{\matOmega}{\bOmega}
\safemath{\matTheta}{\bTheta}

\safemath{\matidentity}{\matI}
\safemath{\matone}{\matO}

\safemath{\rnda}{A}
\safemath{\rndb}{B}
\safemath{\rndc}{C}
\safemath{\rndd}{D}
\safemath{\rnde}{E}
\safemath{\rndf}{F}
\safemath{\rndg}{G}
\safemath{\rndh}{H}
\safemath{\rndi}{I}
\safemath{\rndj}{J}
\safemath{\rndk}{K}
\safemath{\rndl}{L}
\safemath{\rndm}{M}
\safemath{\rndn}{N}
\safemath{\rndo}{O}
\safemath{\rndp}{P}
\safemath{\rndq}{Q}
\safemath{\rndr}{R}
\safemath{\rnds}{S}
\safemath{\rndt}{T}
\safemath{\rndu}{U}
\safemath{\rndv}{V}
\safemath{\rndw}{W}
\safemath{\rndx}{X}
\safemath{\rndy}{Y}
\safemath{\rndz}{Z}

\safemath{\rveca}{\bimA}
\safemath{\rvecb}{\bimB}
\safemath{\rvecc}{\bimC}
\safemath{\rvecd}{\bimD}
\safemath{\rvece}{\bimE}
\safemath{\rvecf}{\bimF}
\safemath{\rvecg}{\bimG}
\safemath{\rvech}{\bimH}
\safemath{\rveci}{\bimI}
\safemath{\rvecj}{\bimJ}
\safemath{\rveck}{\bimK}
\safemath{\rvecl}{\bimL}
\safemath{\rvecm}{\bimM}
\safemath{\rvecn}{\bimN}
\safemath{\rveco}{\bomO}
\safemath{\rvecp}{\bimP}
\safemath{\rvecq}{\bimQ}
\safemath{\rvecr}{\bimR}
\safemath{\rvecs}{\bimS}
\safemath{\rvect}{\bimT}
\safemath{\rvecu}{\bimU}
\safemath{\rvecv}{\bimV}
\safemath{\rvecw}{\bimW}
\safemath{\rvecx}{\bimX}
\safemath{\rvecy}{\bimY}
\safemath{\rvecz}{\bimZ}

\safemath{\rvecxi}{\bmxi}
\safemath{\rveclambda}{\bmlambda}
\safemath{\rvecmu}{\bmmu}
\safemath{\rvectheta}{\bmtheta}
\safemath{\rvecphi}{\bmphi}

\safemath{\rmatA}{\bimA}
\safemath{\rmatB}{\bimB}
\safemath{\rmatC}{\bimC}
\safemath{\rmatD}{\bimD}
\safemath{\rmatE}{\bimE}
\safemath{\rmatF}{\bimF}
\safemath{\rmatG}{\bimG}
\safemath{\rmatH}{\bimH}
\safemath{\rmatI}{\bimI}
\safemath{\rmatJ}{\bimJ}
\safemath{\rmatK}{\bimK}
\safemath{\rmatL}{\bimL}
\safemath{\rmatM}{\bimM}
\safemath{\rmatN}{\bimN}
\safemath{\rmatO}{\bimO}
\safemath{\rmatP}{\bimP}
\safemath{\rmatQ}{\bimQ}
\safemath{\rmatR}{\bimR}
\safemath{\rmatS}{\bimS}
\safemath{\rmatT}{\bimT}
\safemath{\rmatU}{\bimU}
\safemath{\rmatV}{\bimV}
\safemath{\rmatW}{\bimW}
\safemath{\rmatX}{\bimX}
\safemath{\rmatY}{\bimY}
\safemath{\rmatZ}{\bimZ}

\safemath{\rmatDelta}{\bimDelta}
\safemath{\rmatLambda}{\bimLambda}
\safemath{\rmatPhi}{\bimPhi}
\safemath{\rmatSigma}{\bimSigma}
\safemath{\rmatOmega}{\bimOmega}
\safemath{\rmatTheta}{\bimTheta}

\usepackage{amssymb}
\usepackage{amsfonts}
\usepackage{mathrsfs}
\usepackage{xspace}
\usepackage{bm}
\usepackage{fancyref}
\usepackage{textcomp}

\usepackage{multirow}
\usepackage{stmaryrd}

\newenvironment{textbmatrix}{	\setlength{\arraycolsep}{2.5pt}%
								\big[\begin{matrix}}{\end{matrix}\big]%
								\raisebox{0.08ex}{\vphantom{M}}}

\def\be{\begin{equation}}
\def\ee{\end{equation}}
\def\een{\nonumber \end{equation}}
\def\mat{\begin{bmatrix}}
\def\emat{\end{bmatrix}}
\def\btm{\begin{textbmatrix}}
\def\etm{\end{textbmatrix}}

\def\ba#1\ea{\begin{align}#1\end{align}}
\def\bas#1\eas{\begin{align*}#1\end{align*}}
\def\bs#1\es{\begin{split}#1\end{split}}
\def\bg#1\eg{\begin{gather}#1\end{gather}}
\def\bml#1\eml{\begin{multline}#1\end{multline}}
\def\bi#1\ei{\begin{itemize}#1\end{itemize}}

\safemath{\dirac}{\delta}					
\safemath{\krond}{\dirac}					

\safemath{\upto}{\uparrow}
\safemath{\downto}{\downarrow}
\safemath{\iu}{j}							
\safemath{\ev}{\lambda}						
\safemath{\hilseqspace}{l^{2}}				
\newcommand{\banachfunspace}[1]{\setL^{#1}}	
\safemath{\hilfunspace}{\banachfunspace{2}}	

\safemath{\SNR}{\textit{SNR}} 				
\safemath{\PAR}{\textit{PAR}} 				
\safemath{\No}{N_0}							
\safemath{\Es}{E_s}							
\safemath{\Eb}{E_b}							
\safemath{\EbNo}{\frac{\Eb}{\No}}
\safemath{\EsNo}{\frac{\Es}{\No}}

\DeclareMathOperator{\CHop}{\ensuremath{\opH}} 
\safemath{\tvir}{\rndh_{\CHop}}				
\safemath{\tvtf}{\rndl_{\CHop}}				
\safemath{\spf}{\rnds_{\CHop}}				
\safemath{\bff}{H_{\CHop}}					

\safemath{\ircf}{r_{h}}						
\safemath{\tftvcf}{r_{s}}					
\safemath{\tfcf}{r_{l}}						
\safemath{\bfcf}{r_{H}}						

\safemath{\tcorr}{c_h}						
\safemath{\scf}{c_{s}}						
\safemath{\tfcorr}{c_{l}}					
\safemath{\fcorr}{c_{H}}						

\safemath{\mi}{I}							
\safemath{\capacity}{C}						

\safemath{\normal}{\mathcal{N}}			
\safemath{\jpg}{\mathcal{CN}}			
\safemath{\mchain}{\leftrightarrow}		

\safemath{\dB}{\,\mathrm{dB}}
\safemath{\dBm}{\,\mathrm{dBm}}
\safemath{\Hz}{\,\mathrm{Hz}}
\safemath{\kHz}{\,\mathrm{kHz}}
\safemath{\MHz}{\,\mathrm{MHz}}
\safemath{\GHz}{\,\mathrm{GHz}}
\safemath{\s}{\,\mathrm{s}}
\safemath{\ms}{\,\mathrm{ms}}
\safemath{\mus}{\,\mathrm{\text{\textmu}s}}
\safemath{\ns}{\,\mathrm{ns}}
\safemath{\ps}{\,\mathrm{ps}}
\safemath{\meter}{\,\mathrm{m}}
\safemath{\mm}{\,\mathrm{mm}}
\safemath{\cm}{\,\mathrm{cm}}
\safemath{\m}{\,\mathrm{m}}
\safemath{\W}{\,\mathrm{W}}
\safemath{\mW}{\, \mathrm{mW}}
\safemath{\J}{\,\mathrm{J}}
\safemath{\K}{\,\mathrm{K}}
\safemath{\bit}{\,\mathrm{bit}}
\safemath{\nat}{\,\mathrm{nat}}

\safemath{\define}{\triangleq}			

\safemath{\equivalent}{\sim}
\safemath{\distas}{\sim}					
\safemath{\sdiff}{\Delta}				

\safemath{\reals}{\mathbb{R}}
\safemath{\positivereals}{\reals_{+}}
\safemath{\integers}{\mathbb{Z}}
\safemath{\posint}{\integers_{+}}
\safemath{\naturals}{\mathbb{N}}
\safemath{\posnaturals}{\naturals_{+}}
\safemath{\complexset}{\mathbb{C}}
\safemath{\rationals}{\mathbb{Q}}

\newcommand*{\fancyrefapplabelprefix}{app}		
\newcommand*{\fancyrefthmlabelprefix}{thm}		
\newcommand*{\fancyreflemlabelprefix}{lem}		
\newcommand*{\fancyrefcorlabelprefix}{cor}		
\newcommand*{\fancyrefdeflabelprefix}{def}		
\newcommand*{\fancyrefproplabelprefix}{prop}		
\newcommand*{\fancyrefexmpllabelprefix}{exmpl}
\newcommand*{\fancyrefalglabelprefix}{alg}		
\newcommand*{\fancyreftbllabelprefix}{tbl}		

\frefformat{vario}{\fancyrefseclabelprefix}{Section~#1}
\frefformat{vario}{\fancyrefthmlabelprefix}{Theorem~#1}
\frefformat{vario}{\fancyreftbllabelprefix}{Table~#1}
\frefformat{vario}{\fancyreflemlabelprefix}{Lemma~#1}
\frefformat{vario}{\fancyrefcorlabelprefix}{Corollary~#1}
\frefformat{vario}{\fancyrefdeflabelprefix}{Definition~#1}
\frefformat{vario}{\fancyreffiglabelprefix}{Fig.~#1}
\frefformat{vario}{\fancyrefapplabelprefix}{Appendix~#1}
\frefformat{vario}{\fancyrefeqlabelprefix}{(#1)}
\frefformat{vario}{\fancyrefproplabelprefix}{Proposition~#1}
\frefformat{vario}{\fancyrefexmpllabelprefix}{Example~#1}
\frefformat{vario}{\fancyrefalglabelprefix}{Algorithm~#1}

\safemath{\dictab}{[\,\dicta\,\,\dictb\,]}

\safemath{\ysig}{\bmy}
\safemath{\ysighat}{\hat{\ysig}}
\safemath{\ysigdim}{M}
\safemath{\xsig}{\bmx}
\safemath{\xsigdim}{N}
\safemath{\nx}{n_x}
\safemath{\zsig}{\bmz}
\safemath{\zsigdim}{\ysigdim}
\safemath{\rsig}{\bmr}
\safemath{\Adict}{\bA}
\safemath{\Adicttilde}{\widetilde{\Adict}}
\safemath{\Adictdim}{\outputdim\times\xsigdim}
\safemath{\avec}{\bma}
\safemath{\avectilde}{\tilde{\avec}}
\safemath{\Bdict}{\bB}
\safemath{\Bdicttilde}{\widetilde{\Bdict}}
\safemath{\Cdict}{\bC}
\safemath{\cvec}{\bmc}
\safemath{\Ddict}{\bD}
\safemath{\Ddictdim}{\ysigdim\times\xsigdim}
\safemath{\dvec}{\bmd}
\safemath{\Ddicttilde}{\widetilde{\bD}}
\safemath{\Bonb}{\bB}
\safemath{\bvec}{\bmb}
\safemath{\Bonbdim}{\ysigdim\times\ysigdim}
\safemath{\noise}{\bmn}
\safemath{\noisedim}{\ysigim}
\safemath{\err}{\bme}
\safemath{\errdim}{\ysigdim}
\safemath{\errset}{\setE}
\safemath{\nerr}{n_e}
\safemath{\delop}{\bP_\errset}
\safemath{\delopc}{\bP_{{\errset}^c}}

\safemath{\cplxi}{\imath}
\safemath{\cplxj}{\jmath}

\safemath{\dict}{\matD}
\safemath{\inputdim}{N}		
\safemath{\outputdim}{M}		
\safemath{\sparsity}{S}	
\safemath{\inputdimA}{{N_a}}	
\safemath{\inputdimB}{{N_b}}	
\safemath{\elemA}{{n_a}}	
\safemath{\elemB}{{n_b}}	
\safemath{\resA}{\matR_a}	
\safemath{\resB}{\matR_b}	
\safemath{\subD}{\matS} 
\safemath{\subA}{\matS_a} 
\safemath{\subB}{\matS_b} 
\safemath{\dicta}{\matA} 	
\safemath{\dictb}{\matB} 	
\safemath{\hollowS}{H}
\safemath{\hollowA}{H_a}
\safemath{\hollowB}{H_b}
\safemath{\cross}{Z}
\safemath{\coh}{\mu_d}			
\safemath{\coha}{\mu_a}			
\safemath{\cohb}{\mu_b}			
\safemath{\mubs}{\nu}	
\safemath{\cohm}{\mu_m} 
\safemath{\dictset}{\setD}	
\safemath{\dictsetp}{\dictset(\coh,\coha,\cohb)}	
\safemath{\dictsetgen}{\dictset_\text{gen}}
\safemath{\dictsetgenp}{\dictsetgen(\coh)}
\safemath{\dictsetonb}{\dictset_\text{onb}}
\safemath{\dictsetonbp}{\dictsetonb(\coh)}

\safemath{\leftside}{U}
\safemath{\rightsideA}{R_a}
\safemath{\rightsideB}{R_b}

\safemath{\indexS}{\setI_S} 

\safemath{\na}{n_a}			
\safemath{\nb}{n_b}			
\safemath{\coeffa}{p_i}	
\safemath{\coeffb}{q_j}	
\safemath{\seta}{\setP}		
\safemath{\setb}{\setQ}     
\safemath{\setw}{\setW}	
\safemath{\setz}{\setZ}	
\safemath{\cola}{\veca}		
\safemath{\colb}{\vecb}		
\safemath{\cold}{\vecd}		
\safemath{\inputvec}{\vecx} 	
\safemath{\error}{\vece}	
\safemath{\noiseout}{\vecz} 	
\safemath{\inputvecel}{x}
\safemath{\inputveca}{\vecx_a}
\safemath{\inputvecb}{\vecx_b}
\safemath{\outputvec}{\vecy}	
\safemath{\lambdamin}{\lambda_{\mathrm{min}}}

\safemath{\elltwo}{\ell_2}
\safemath{\ellone}{\ell_1}
\safemath{\ellzero}{\ell_0}
\safemath{\ellinf}{\ell_\infty}
\safemath{\ellinftilde}{\ell_{\widetilde\infty}}
\safemath{\licard}{Z(\coh,\coha,\cohb)}
\safemath{\xsol}{\hat{x}}
\safemath{\xbord}{x_b}		
\safemath{\xstat}{x_s}		
\safemath{\xstatLone}{\tilde{x}_s}
\safemath{\order}{\mathcal{O}} 
\safemath{\scales}{\Theta} 
\safemath{\ones}{\mathbf{1}} 
\safemath{\zeroes}{\mathbf{0}} 
\safemath{\thlone}{\kappa(\coh,\cohb)} 
\safemath{\constoneA}{\delta} 
\safemath{\constoneB}{\epsilon} 
\safemath{\nlarge}{L}				   
\safemath{\sumlarge}{S_\nlarge}
\safemath{\maxlarger}{P_\nlarge}	   
\safemath{\Pzero}{\textrm{P0}}	
\safemath{\Pone}{\textrm{P1}}
\safemath{\vecfir}{\vecw}			 
\safemath{\vecsec}{\vecz}
\safemath{\elvecfir}{w}              
\safemath{\elvecsec}{z}				 
\safemath{\nlargefir}{n}
\safemath{\normout}{\gamma}
\safemath{\auxfun}{h}
\safemath{\supp}{\textrm{supp}}

\safemath{\indexa}{\ell}
\safemath{\indexb}{r}
\safemath{\indexc}{i}
\safemath{\indexd}{j}

\safemath{\project}{P}

\makeatletter
\def\@startsection#1#2#3#4#5#6{\if@noskipsec \leavevmode \fi
   \par \@tempskipa #4\relax
   \@afterindenttrue
   \ifdim \@tempskipa <\z@ \@tempskipa -\@tempskipa \@afterindentfalse\fi
   \if@nobreak \everypar{}\else
     \addpenalty{\@secpenalty}\addvspace{\@tempskipa}\fi \@ifstar
     {\@dblarg{\@sect{#1}{#2}{#3}{#4}{#5}{#6}}}%
     {\@dblarg{\@sect{#1}{#2}{#3}{#4}{#5}{#6}}}}

\def\section{\@startsection {section}{1}{\z@}{20pt plus 2pt minus 2pt}
{8pt plus 2pt minus 2pt}{\centering\normalsize\bf}}

\def\subsection{\@startsection {subsection}{2}{\z@}{16pt plus 2pt minus 2pt}
{6pt plus 2pt minus 2pt}{\normalsize\it}}

\long\def\@makecaption#1#2{
\vskip10pt\begin{center} #1 #2 \end{center}\par\vskip 1pt}
\def\fnum@figure{\raggedright{\small Fig. \thefigure }.%
\small}
\def\fnum@table{\small TABLE \thetable\\\small}
\def\thetable{\Roman{table}}

\def\thebibliography#1{\section*{References\@mkboth
 {REFERENCES}{REFERENCES}}\list
 {[\arabic{enumi}]}{\settowidth\labelwidth{[#1]}\leftmargin\labelwidth
 \parsep 0pt \itemsep 1pt plus 2pt minus 2pt
 \advance\leftmargin\labelsep
 \usecounter{enumi}}
 \def\newblock{\hskip .11em plus .33em minus .07em}
 \sloppy\clubpenalty4000\widowpenalty4000
 \sfcode`\.=1000\relax}

\makeatother

\usepackage{color}

\definecolor{gray}{HTML}{555555}   
\definecolor{colcirc}{RGB}{0,0,71}

\newcommand*\tinygraycircled[1]{\Circled[inner color=white, fill color= gray, outer color=gray]{\scriptsize{#1}}}

\begin{document}

\date{}

\title{\Large\bf				       
A Jammer-Mitigating 267\,Mb/s 3.78\,mm$^\textbf{2}$ 583\,mW\\ 32$\times$8 Multi-User MIMO Receiver
in 22FDX\\[-0.3cm]
}

\author{{\large Florian Bucheli, Oscar Casta\~neda, Gian Marti, and Christoph Studer}  \\
{\normalsize Dept. of Information Technology and Electrical Engineering, ETH Zurich, Switzerland; email: studer@ethz.ch}  
\\[-0.5cm]
}

\maketitle

\thispagestyle{empty}

\section{Abstract\\[-1.5ex]}
We present the first multi-user (MU) multiple-input multiple-output (MIMO) 
receiver ASIC that mitigates jamming attacks.
The ASIC implements a recent nonlinear algorithm that performs joint 
jammer mitigation (via spatial filtering) and data detection 
(using a box prior on the data symbols).
Our design supports 8 user equipments (UEs) and 32 basestation (BS) antennas, 
QPSK and 16-QAM with soft-outputs, 
and enables the mitigation of single-antenna barrage jammers \emph{and} smart jammers.
The fabricated $22$\,nm FD-SOI ASIC includes preprocessing,
has a core area of $3.78$\,mm$^2$,
achieves a throughput~of $267$\,Mb/s while consuming $583$\,mW,
and is the only existing design that enables reliable data detection
under jamming attacks.

\vspace{-0.5cm}

\section{Introduction\\[-1.5ex]}
Jammer-resistant wireless systems are urgently needed in a world where critical infrastructure increasingly
depends on wireless communications. 
MIMO technology enables jammer mitigation through spatial filtering~\cite{Pirayesh1}. 
For the mitigation of smart jammers
that only jam certain signal parts (e.g., the pilot or data phase), 
such spatial filters must rely on nonlinear methods~\cite{Marti1}.  
Moreover, these spatial filters need to be implemented in hardware to support 
the high throughputs required in modern wireless systems.
To this day, however, there are \emph{no} hardware implementations of such jammer-resistant wireless receivers.

We present the first jammer-mitigating \mbox{MU-MIMO} receiver ASIC. 
Our ASIC implements a variant of the nonlinear algorithm SANDMAN from \cite{Marti1} 
to enable the mitigation of barrage jammers \emph{and} smart jammers against which linear methods fail. 
SANDMAN processes the receive data in blocks that consist of multiple pilot and data symbols; 
see \fref{fig:sysmodel} for the system model. 
Using a linear estimate of the UEs' channel matrix, SANDMAN approximately solves 
the non-convex optimization problem outlined in \fref{fig:signal_proc} to estimate and null the jammer subspace, 
while simultaneously detecting the UE transmit data of the entire block.
SANDMAN achieves this by using an iterative algorithm based on projected gradient descent, shown in~\fref{fig:flow_chart}.

Our ASIC supports a $32\times8$ MU-MIMO system, where $32$ is the number of BS antennas
and $8$ is the number of simultaneously transmitting UEs. The data is processed in blocks of $64$ symbols
($16$ pilot symbols and $48$ data symbols). The single-antenna jammer may jam continuously (barrage), 
or only during the pilot symbols, or only during the data symbols. 
The ASIC performs channel estimation (CHEST) and produces soft-outputs for the data bits that can be  used for decoding. 
QPSK and 16-QAM are supported as transmit constellations. 	  
\vspace{-0.5cm}

\section{Architecture and Operation\\[-1.ex]}
\fref{fig:flow_chart} shows the SANDMAN algorithm~\cite{Marti1} for single antenna jammers, reorganized into a hardware-efficient schedule.
This implementation of SANDMAN involves complex-valued matrix-matrix products (\tinygraycircled{1},~\tinygraycircled{7}) and additions/subtractions (\tinygraycircled{1},~\tinygraycircled{6},~\tinygraycircled{8}), matrix-vector products (\tinygraycircled{2},~\tinygraycircled{3},~\tinygraycircled{5}), and scalar operations (\tinygraycircled{4},~\tinygraycircled{6}, \tinygraycircled{8}).
To minimize latency, we employ the highly parallel spatial architecture illustrated in \fref{fig:ingredient_overview}.
The architecture is mainly composed of a 2D array of $32\times8$ processing elements (PEs), where each PE consists of a highly reconfigurable multiply-accumulate (MAC) unit, a ``prox'' unit for thresholding, and flip-flop (FF) arrays for local storage.
The size of the 2D array was chosen to match that of the channel matrix estimate~$\textstyle\hat\bH$ to facilitate the computationally demanding matrix multiplications involving $\textstyle\hat\bH$~(\tinygraycircled{1}) and $\textstyle\hat\bH^\text{H}$~(\tinygraycircled{7}).
The $32\times8$ array is divided into four $8\times8$ PE slices connected column-wise. 
Each PE slice enables (through circular shifting of the input values) the application of Cannon's algorithm for multiplication by an $8\times8$ matrix and subsequently by its Hermitian, without having to explicitly transpose the matrix in memory. 
Matrix multiplications involving larger outer dimensions (as in \tinygraycircled{1}, \tinygraycircled{7}) are executed and stored in $8\times8$-sized blocks.
To engage all PEs simultaneously when computing matrix-vector products (\tinygraycircled{2}, \tinygraycircled{3}, \tinygraycircled{5}), entries can be broadcasted to a whole column or row.
Since these matrix-vector products have an inner dimension larger than $8$, the partial sums of the $8\times8$ PE slices need to be summed up row- or column-wise. For this, we support the reconfiguration of PEs on a row/column into row/column adders (see bottom of \fref{fig:ingredient_overview}).
Jammer renormalization (\tinygraycircled{4}) requires high numerical precision due to the jammer's large dynamic range. To preserve energy efficiency, we use a specialized set of extended PEs (PE+) with larger bitwidths for this step. The renormalization also uses a look-up-table-based inverse square root unit. 
We implement the pseudorandom vector \bmx (\tinygraycircled{2}) with a circular shift register. 
After $t_\text{max}\!=\!10$ algorithm iterations, a soft-output/LLR unit converts the two dimensions of a complex-valued scalar symbol estimate into two/four LLRs. These LLRs correspond to the two/four bits of a Gray-mapped QPSK/16-QAM constellation with a box size of~$\sfrac{1}{\sqrt{2}}$. The resulting LLRs are stored in the S FF array (see \fref{fig:ingredient_overview}). \fref{fig:phases} shows how the PEs are configured at the start of each operation (\tinygraycircled{1}-\tinygraycircled{8}) and which hardware elements are used in that operation.
\vspace{-0.5cm}

\section{Results and Conclusion\\[-1.ex]}
\fref{fig:BER} shows the bit error-rate (BER) vs. SNR performance of different jammers and different constellations for the hardware implementation of SANDMAN compared to linear minimum-mean squared-error (LMMSE), channel hardening-exploiting message passing (CHEMP)~\cite{Naras1}, and approximate message passing (LAMA)~\cite{Jeon1}. 
All jammers in \fref{fig:BER} transmit at $30$\,dB higher receive power than the average UE.
SANDMAN significantly outperforms the other receivers for all jammers/constellations since it is the only implemented MU-MIMO receiver that can mitigate jammers (and in particular, smart jammers).
The fixed-point performance of the SANDMAN ASIC follows closely the performance of double-precision floating point (see~\fref{fig:BER}).

\fref{fig:chip} shows a micrograph of the ASIC fabricated in $22$\,nm FD-SOI with the main blocks highlighted. 
The core occupies $3.78\,\text{mm}^\text{2}$ of a total chip area of $5\,\text{mm}^\text{2}$\!. 
\fref{fig:measurements} shows voltage-frequency scaling results for different body biases.
As expected, the maximum clock frequency increases with the core supply, until at some point it drops rapidly due to signal integrity issues related to the used QFN56 package.
At $0.78$\,V core supply, zero body biasing, $300$\,K, 16-QAM and $t_\text{max}\!=\!10$ algorithm iterations, the ASIC reaches a clock frequency of $320$\,MHz and a throughput of $267$\,Mb/s while consuming $583$\,mW.

\fref{tbl:comparison} compares our ASIC to state-of-the-art MU-MIMO detectors.
Our design is the only one that enables data detection under jamming attacks; all other designs would completely fail in that scenario, see Fig. 5. This jammer resilience comes at the cost of lower throughput and efficiency, a price worth paying in mission-critical applications where a robust link is paramount.
\vspace{-0.5cm}

\section{Acknowledgments\\[-1ex]}
The authors would like to thank GlobalFoundries for providing silicon fabrication through the 22FDX university program.
\vspace{-0.8cm}

\vspace{-0.15cm}


\newpage

\onecolumn

\noindent\begin{minipage}{0.25\textwidth}
\centering
\begin{figure}[H]
\centering
\includegraphics[height=4.1cm]{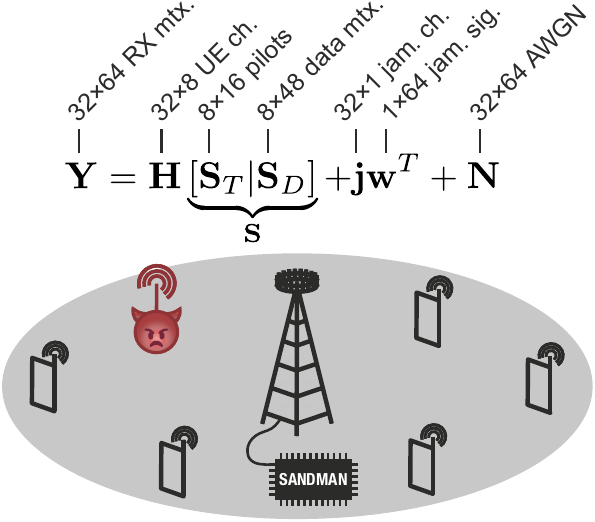}
\vspace{-0.55cm}
\caption{System model of the jammed\!\! multi-user (MU) MIMO uplink.}
\label{fig:sysmodel}
\end{figure}
\end{minipage}
\hspace{1mm}
%
\noindent\begin{minipage}{0.3\textwidth}
\centering
\begin{figure}[H]
\centering
\includegraphics[height=3.9cm]{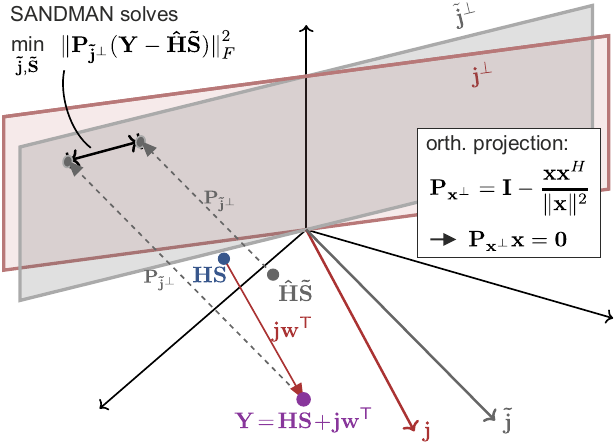}
\vspace{-0.35cm}
\caption{SANDMAN approximately solves a non-convex optimization problem.}
\label{fig:signal_proc}
\end{figure}
\end{minipage}
\hspace{1mm}
\noindent\begin{minipage}{0.40\textwidth}
\centering
\begin{figure}[H]
\includegraphics[height=4.3cm]{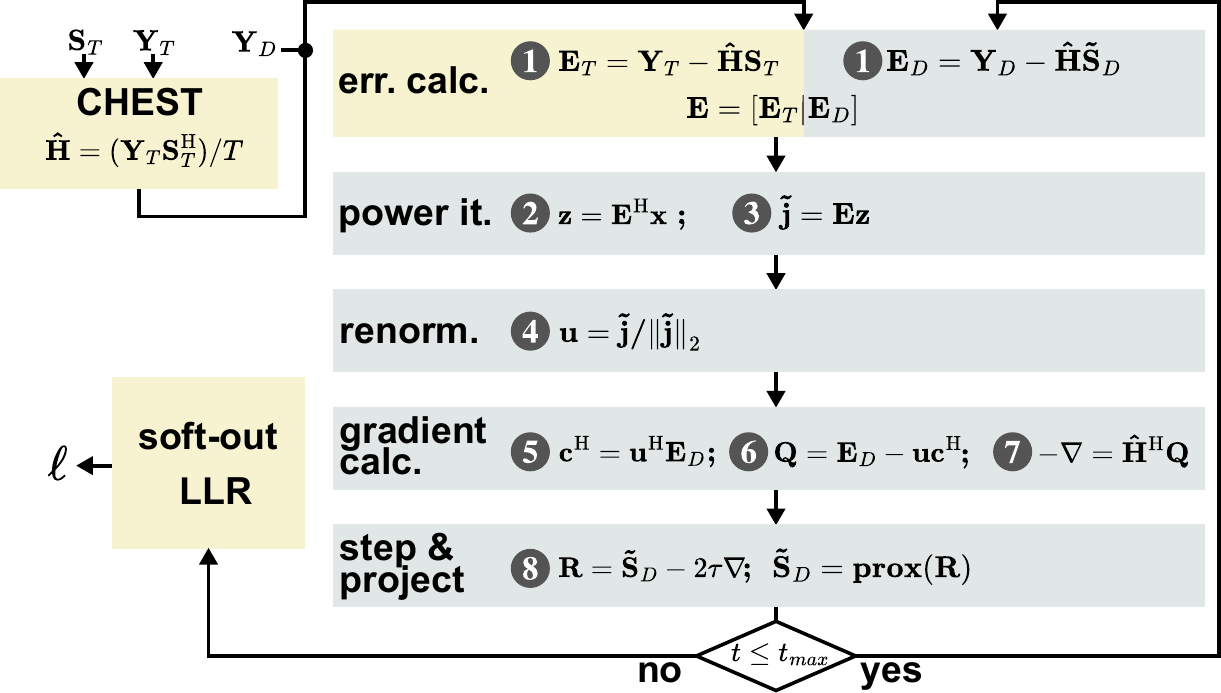}
\vspace{-0.74cm}
\caption{Outline of the SANDMAN algorithm, which runs for $t_\text{max}$ iterations. Steps only executed once are in yellow.}
\label{fig:flow_chart}
\end{figure}
\end{minipage}

\vspace{-0.5cm}
\begin{figure}[H]
\begin{minipage}[h]{0.53\textwidth}
\begin{center}
\includegraphics[width=\linewidth]{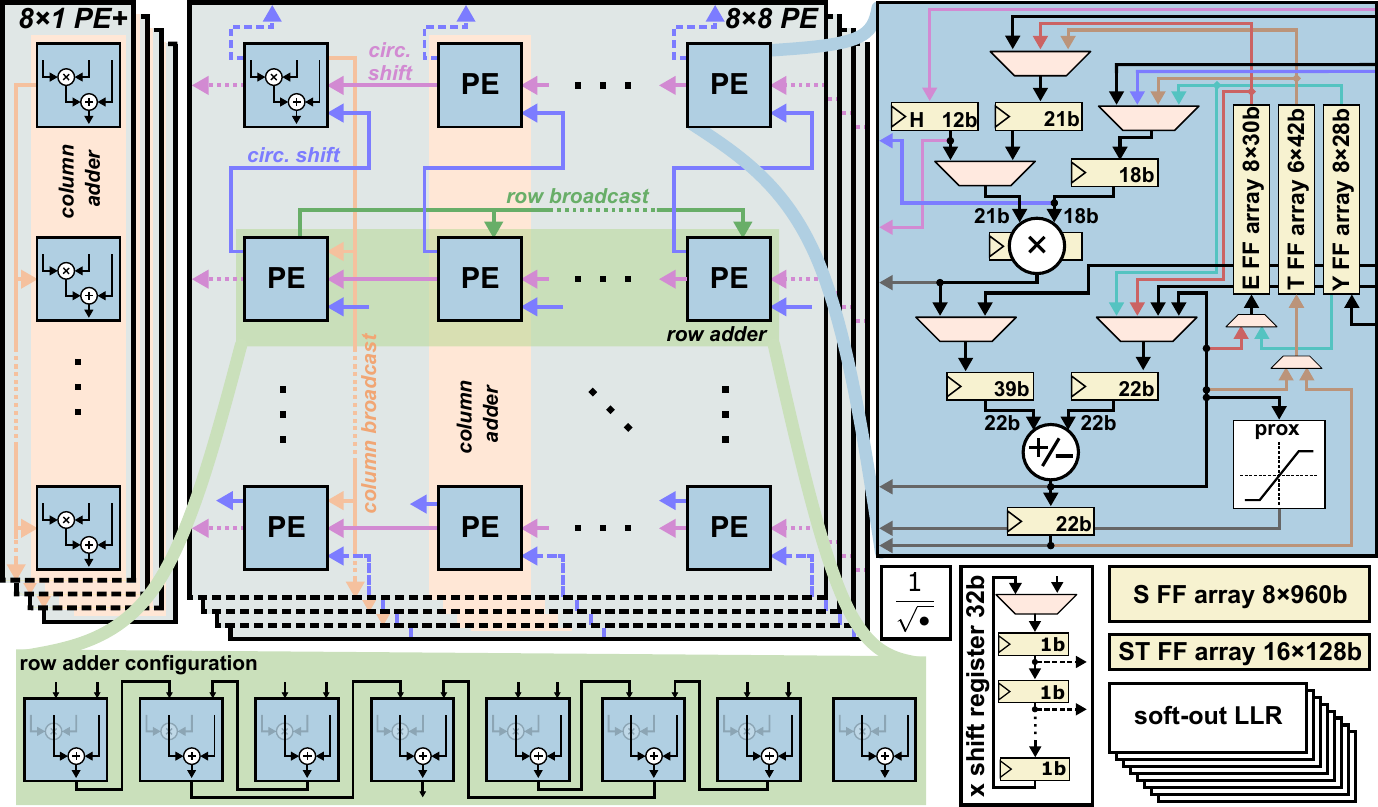} 
\vspace{-5mm}
\caption{Overview of the SANDMAN architecture and implementation details of the
processing elements (PEs). The control module is not shown. Rows and
columns of PEs can be reconfigured into row adders and column adders.}
\label{fig:ingredient_overview}
\end{center} 
\end{minipage}
\\
%
%
\noindent\begin{minipage}[t]{0.55\textwidth}
\vspace{-0.5cm}
\begin{figure}[H]
\includegraphics[width=0.985\linewidth]{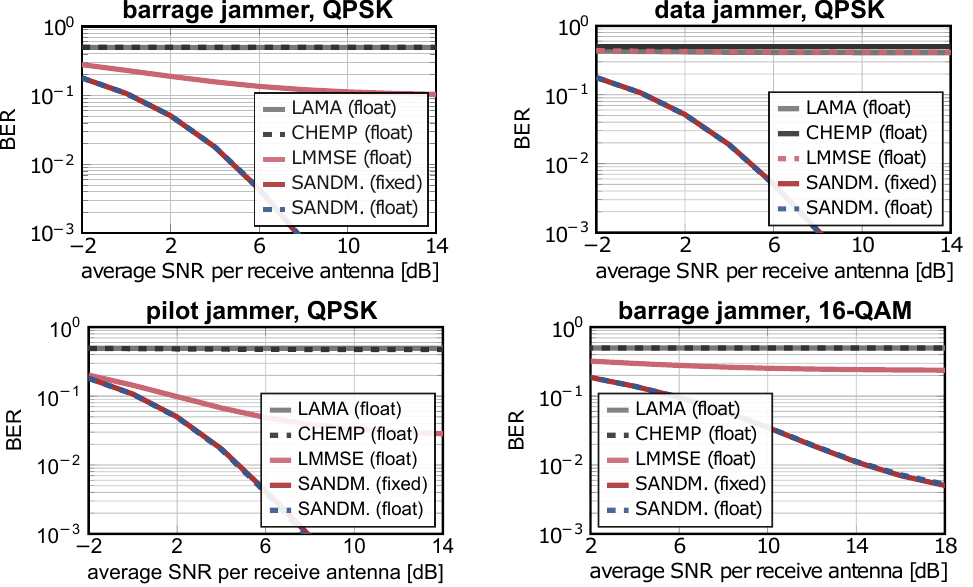}
\vspace{-0.25cm}
\caption{Performance for $32\times8$ MIMO with different jammers/constellations.
}
\label{fig:BER}
\end{figure}
\end{minipage}
\noindent\begin{minipage}[t]{0.445\textwidth}
\vspace{-7.56cm}
\centering
\begin{figure}[H]
\vspace{-3mm}
\centering
\hfill
\includegraphics[width=\linewidth]{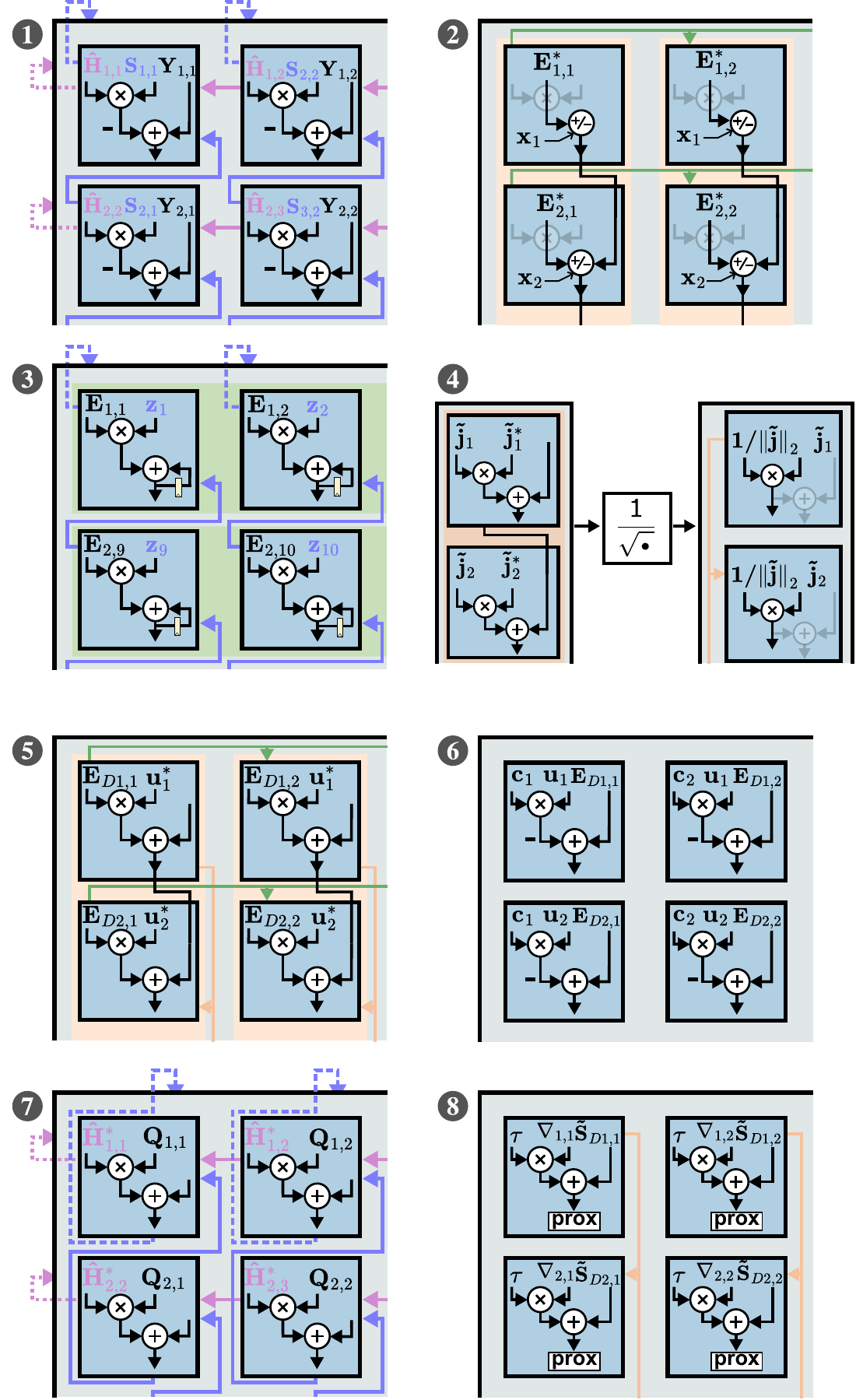}
\vspace{-0.495cm}
\caption{Different configurations of the PE array used to execute the operations
 in \fref{fig:flow_chart}; operation \tinygraycircled{4} relies on the PE+ array.}
\vspace{-0.5cm}
\label{fig:phases}
\end{figure}
\end{minipage}

\end{figure}

\vspace{1.8cm}

\noindent\begin{minipage}{0.26\textwidth}
\centering
\begin{figure}[H]
\vspace{2mm}
\includegraphics[height=5.1cm]{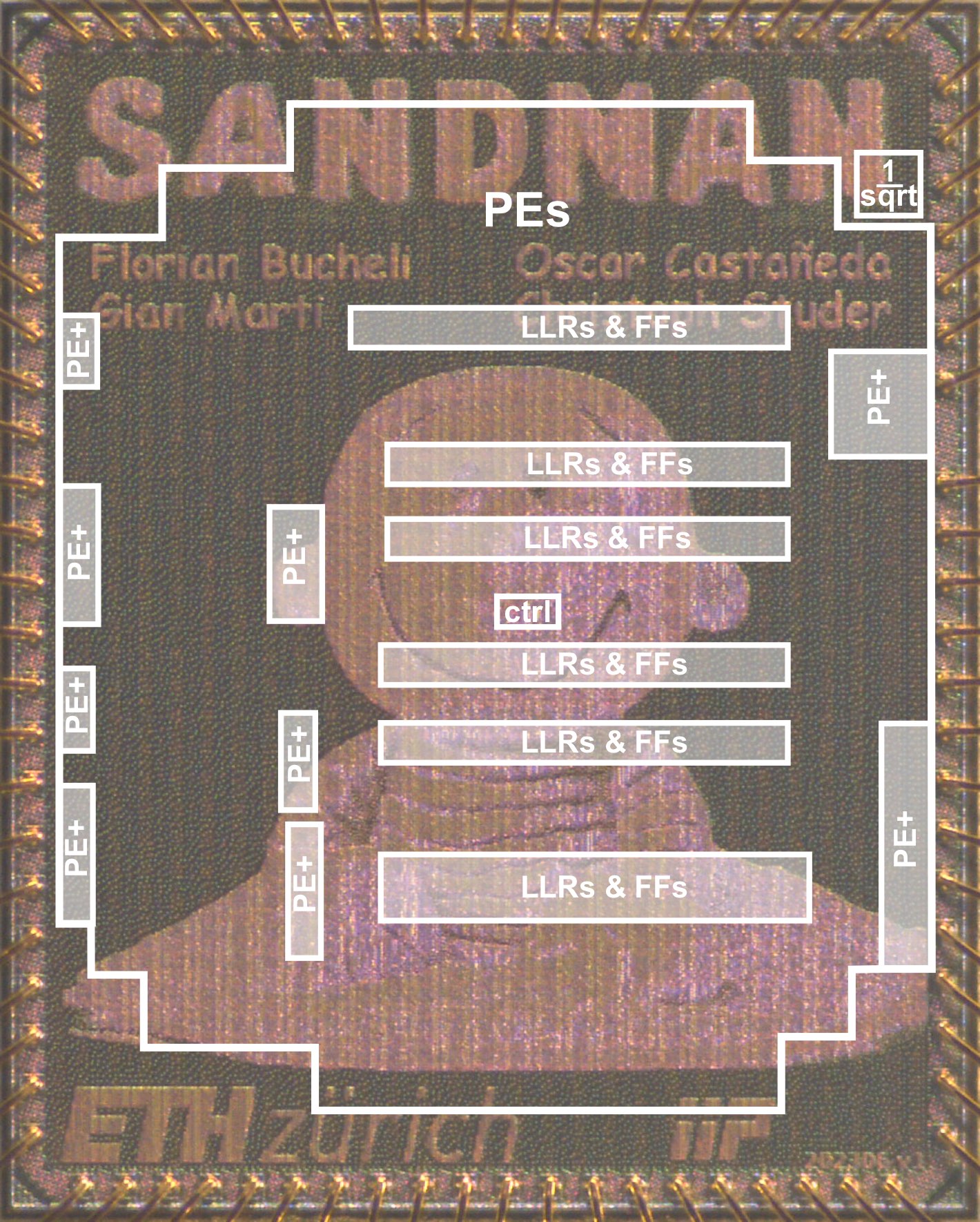}
\vspace{-3mm}
\caption{Chip micrograph with highlighted modules.}
\vspace{-2mm}
\label{fig:chip}
\end{figure}
\end{minipage}
\hspace{-5mm}
\hfill
\noindent\begin{minipage}{0.26\textwidth}
\centering
\begin{figure}[H]
\vspace{1mm}
\hspace{1mm}
\includegraphics[height=5.1cm]{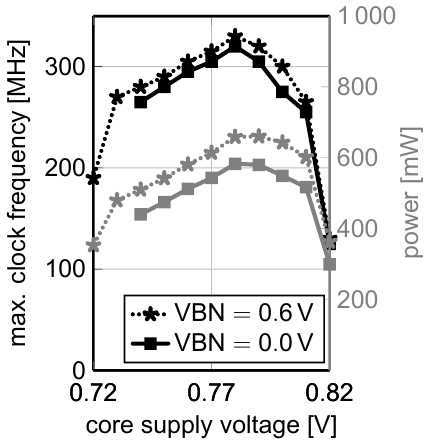}
\vspace{-5.4mm}
\caption{\mbox{Measured frequency (black) and\hspace{-4mm}} \mbox{power (gray) vs. core and body voltages.}}
\vspace{-2mm}
\label{fig:measurements}
\end{figure}
\end{minipage}
\hfill
\noindent\begin{minipage}{0.45\textwidth}
\begin{center}  
\vspace{0.3cm}
\captionof{table}{Measurement results and ASIC comparison}
\label{tbl:comparison}
\scalebox{0.8}{
\renewcommand{\arraystretch}{1.14}     
\footnotesize
\begin{tabular}{@{} l c | c c c @{}} 
\toprule[0.2em]
\multirow{2}{4em}{}
& 
This work\!  & Jeon \cite{Jeon1} & Tang \cite{Tang1} & Prabhu~\cite{Prabhu1} \\
\midrule[0.15em]
Max. BS antennas	 	& 32 & 256  & 128 & 128 \\ 
Max. UEs  			& 8 & 32  & 32 & 8 \\ 
Modulation [QAM]\!\!\!\!	&  16 &  256 &  256  &  256 \\
Soft-outputs		& yes  & yes & yes  & no \\
Algorithm 
					& SANDMAN & LAMA & CHEMP & LMMSE \\
CHEST~\& preproc. incl.\!\!\!\! 
					& {\bf yes} & no & no & {\bf yes} \\
Jammer mitigation\!\!
 					& {\bf yes}  & {\bf no} & {\bf no} & {\bf no}\\ 
\midrule[0.15em]
Technology [nm] 	& 22  & 28 & 40 & 28 \\
Core supply [V]        & 0.78 & 0.9 & 0.9 & 0.9 \\
Core area~[$\text{mm}^2$] 	& 3.78 & 0.37 & 0.58 & $-$ \\
Max. frequency~[MHz] 			& 320 & 400 & 425 & 300 \\
Power~[mW] 				& 583 & 151 & 221 & 18 \\
Throughput~[Mb/s]\!\! 
						& 267 & 354 & 2\,760 & 300 \\ 
Area~eff.$^a$~[Mb/s/$\text{mm}^2$]\!\!\!\!\! 
						&  70 & 1\,974 & 28\,602 & $-$ \\
Energy$^a$~[pJ/b]		& 2\,187 & 252 & 33 & 35 \\
\bottomrule[0.2em]
\end{tabular}}\\
\scriptsize
\hspace{-1.2cm}
$^a$Technology normalized to 22\,nm and 0.78\,V core supply.
\end{center}
\vspace{-5mm}
\end{minipage}

\thispagestyle{empty}

\end{document}